\documentclass{aa}
\usepackage{graphics}
\begin{document}

\thesaurus{11.02.1; 11.02.2; 13.25.2}

\title{The 0.1--200 keV spectrum of the blazar PKS 2005--489
during an active state}

\author{G. Tagliaferri\inst{1} 
\and G. Ghisellini\inst{1} 
\and P. Giommi\inst{2}
\and A. Celotti\inst{3}
\and M. Chiaberge\inst{3}
\and L. Chiappetti\inst{4}
\and I.S. Glass\inst{5}
\and L. Maraschi\inst{6}
\and F. Tavecchio\inst{6} 
\and A. Treves\inst{7}
\and A. Wolter\inst{6}
}

\offprints{G. Tagliaferri}

\institute{Osservatorio Astronomico di Brera, Via Bianchi 46, I--23807 Merate,
Italy
\and {\it Beppo}SAX Science Data Center, ASI, Via Corcolle, 19, 
          I--00131 Roma, Italy
\and SISSA/ISAS, via Beirut 2--4, 34014 Trieste, Italy
\and Istituto di Fisica Cosmica G.Occhialini, CNR, Via Bassini 15, 
     I-20133 Milano, Italy
\and South African Astronomical Observatory, PO Box 9, Observatory 7935, South Africa
\and Osservatorio Astronomico di Brera, Via Brera, 28, I-20121 Milano, Italy
\and Istituto di Fisica, Universit\`a dell'Insubria, Via Lucini 3, I--22100 
     Como, Italy
}

\date{Received date / accepted date}

\titlerunning{The 0.1--200 keV spectrum of PKS 2005--489}
\authorrunning{Tagliaferri et al.\ }

\maketitle

\begin{abstract}

{
The bright BL Lac object PKS 2005--489
was observed by $Beppo$SAX on November 1--2, 1998,
following an active X--ray state detected by {\it Rossi}XTE.
The source, detected between 0.1 and 200 keV, was in a very high
state with a continuum well fitted by a steepening spectrum due to 
synchrotron emission only.
Our X--ray spectrum is the flattest ever observed for this source.
The different X--ray spectral slopes and fluxes, as measured by
various satellites, are consistent with relatively little 
changes of the  peak frequency of the synchrotron emission, 
always located below  $10^{17}$ Hz. We discuss these results 
in the framework of synchrotron self--Compton models. 
We found that for the $Beppo$SAX observation, the synchrotron 
peak frequency is between $10^{15}$ and $2.5 \times 10^{16}$ Hz,
depending on the model assumptions.
}
                
\end{abstract}

\section{Introduction}

The overall spectral energy distribution (SED) of blazars
(BL Lac objects and violently variable quasars) shows 
(in a $\nu$ vs $\nu F_{\nu}$ representation) two broad 
emission peaks: the lower frequency peak is believed to be 
produced by synchrotron emission, while the higher frequency 
one is probably due to the inverse Compton process.
The location of the synchrotron peak is used to define different 
classes of blazars: HBL (High frequency peaked blazar, with the
maximum power output in the UV or X--ray frequencies) and LBL (Low
frequency peaked blazar, peaking in the IR or optical bands)
(Giommi \& Padovani 1994). Thus, for LBL sources the X--ray 
emission is dominated by the inverse Compton component.

The continuum emission of blazars is both highly luminous and rapidly 
variable and it is probably dominated by emission from a jet
moving  relativistically at small angles to the line
of sight (Blandford \& Rees 1978). Strong evidence for bulk
relativistic motion was first provided by multiwavelength studies, 
later confirmed directly with VLBI observations (Vermeulen \& Cohen 1994).
However, single epoch spectra cannot constrain the models of variability
in relativistic jets (e.g. K\"onigl 1989; Ulrich et al. 1997).
Since blazars emit over the entire electromagnetic spectrum, a key for 
understanding blazar variability is the acquisition of several wide
band spectra in different luminosity states during major flaring episodes.
Coupling spectral and temporal information greatly constrains the jet
physics, since different models predict different variability as a
function of wavelength.
Important progress in this respect has been achieved recently for
some of the brightest and best studied blazars, as PKS 2155--304 
(Chiappetti et al. 1999; Urry et al. 1997), BL Lac (Bloom et al. 1997),
3C 279 (Wehrle et al. 1998), Mkn 501 (Pian et al. 1998), Mkn 421
(Maraschi et al. 1999; Fossati et al. 2000a,b), ON 231 
(Tagliaferri et al. 2000).
We successfully used the {\it Beppo}SAX satellite (Boella et al. 1997a)
to perform observations of blazars that were known to be in a high state
from observations carried out both in other bands (mainly optical and TeV)
and in the X--ray band itself.
The good {\it Beppo}SAX sensitivity and spectral resolution over
a very wide X--ray energy range are ideal to constrain the existing 
models for the X--ray emission.

PKS 2005--489 ($z=0.071$) is a bright BL Lac object
that belongs to the 1 Jy radio catalog (Stickel et al. 1991).
Its broad band spectrum peaks in the UV-- soft X--ray bands
making it an HBL type of source. 
It is one of the few extragalactic objects seen in the EUV band 
(Marshall et al. 1995) and it has been detected in the $\gamma$--ray band by
EGRET (von Montigny et al. 1995). 
In the X--ray band it has been observed 5 times with EXOSAT,
showing large flux variations correlated with changes in the 
spectrum (harder spectrum with increasing flux) (Giommi et al. 1990, 
Sambruna et al. 1994a,b). These results are also confirmed by two ROSAT 
observations, which also show a steep 0.2-2.0 keV spectrum, with
photon index of about 3 (Sambruna et al. 1995). In September 1996, 
PKS 2005--489 has been observed by {\it Beppo}SAX,
that detected the source in a rather bright state, with a flatter 
2--10 keV photon index of 2.3 (Padovani et al. 1988, 2000).
Finally, a monitoring campaign has been performed in the period 
October--December 1998 with
{\it Rossi}XTE and the source has been detected up to 40 keV
(Perlman et al. 1999). During this campaign the source underwent
a strong flare, with a peak to quiescent 2-10 keV flux variation
of a factor of 8 (30 with respect to the ROSAT fluxes). 
This event
essentially lasted for all the {\it Rossi}XTE campaign, peaking
on 10 November, thus the source was in a high state for at least 
three months (Perlman et al. 1999). During the week of 18--25 October,
a flare alert was issued by the all-sky monitor (ASM) team
(Remillard 1998). This triggered our ToO observation and
PKS 2005--489 was observed with {\it Beppo}SAX in November 1--2, 1998.

In this paper we present and discuss the results of this $Beppo$SAX
observation together with some simultaneous infrared data.

\section{X--ray Observations}

\subsection{Observations and Data Reduction}

\begin{figure}
\begin{center}
{\resizebox{\hsize}{!}{\includegraphics{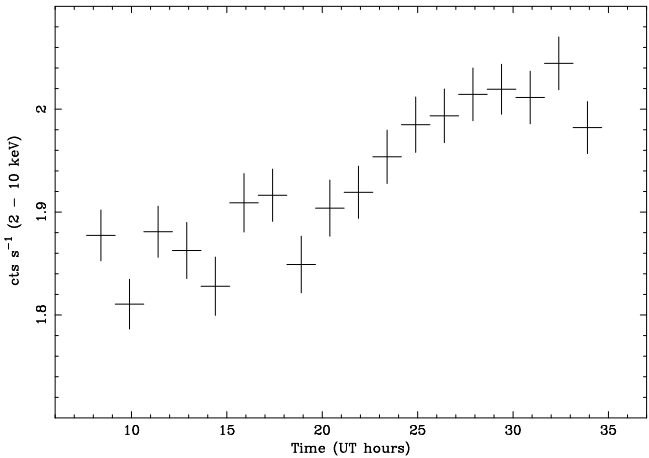}}}\\
\end{center}
\caption{
The 2--10 keV light curve of PKS\,2005--489 during the {\it Beppo}SAX 
observation of 1-2 November, 1998. The bin time is of 5400\,s.
}
\end{figure}

The {\it Beppo}SAX satellite is an Italian mission with the
participation of the Netherlands
Agency for Aerospace Programs (NIVR) and the Space Science Department
of the European Space Agency (SSD-ESA). It carries on board four Narrow
Field Instruments (NFI) pointing in the same direction and covering a 
very large energy range from 0.1 to 300 keV (Boella et al. 1997a).
Two of the four instruments have imaging capability, the Low Energy
Concentrator Spectrometer (LECS), sensitive in the range 0.1--10 keV
(Parmar et al. 1997), and the three Medium Energy Concentrator
Spectrometers  (MECS) sensitive in the range 1.3--10 keV (Boella
et al. 1997b). The LECS and three MECS detectors are all Gas
Scintillation Proportional Counters and are at the focus of four identical
grazing incidence X--ray telescopes. The other two are passively collimated
instruments: the High Pressure Proportional Counter (HPGSPC), sensitive
in the range 4--120 keV (Manzo et al. 1997) and the Phoswich Detector
System (PDS), sensitive in the range 13--300 keV (Frontera et al. 1997). 
For a full description of the {\it Beppo}SAX mission see 
Boella et al. (1997a).

The data analysis for the LECS and MECS instruments was based on the 
linearized, cleaned event files obtained from the online archive 
(Giommi \& Fiore 1998). Light curves and spectra were accumulated 
with the FTOOLS package (v. 4.0), using an extraction 
region of 8.5 and 4\,arcmin radius for the LECS and MECS, respectively.
At low energies the LECS has a broader Point Spread Function (PSF) than the 
MECS, while above 2\,keV the PSFs are similar. The adopted regions provide 
more than 90\%
of the source counts at all energies both for the LECS and MECS. 

As expected, PKS 2005--489 was detected in a high state
in the X--ray band. However, our observation occurred about one week
before the source reached its maximum as monitored 
by {\it Rossi}XTE (Perlman et al. 1999).
Some flux variability of the order of 10\%
on a timescale of about 10 hours
was detected during our observations in the 2--10 kev energy band
(see Fig.\,1). No similar flux variability was detected below 1 keV.
Given that the amount of variability is quite small, we extracted only one
spectrum for the full observation. This corresponds to a total LECS exposure 
time of 20395 s and MECS  exposure time of 52478 s.
The background files for the spectral analysis were accumulated
from the long blank field exposures available from the SDC public ftp site
(see Fiore et al. 1999, Parmar et al. 1999).

The HPGSPC and PDS were operated in the customary collimator rocking mode,
with each collimator pointing alternatively at the source and at the
background for 96 s. The HPSGPC has a single collimator, thus the whole
instrument is looking at the target for half of the time and to an offset
direction for the other half. The total exposure time in our case was 23258 s.
The PDS has two collimators, thus one is always looking at the source and
the other one at the background. The total PDS exposure time, for the
two collimators, was of about 46.8 ks.
The background subtracted HPGSPC and PDS spectra were obtained from
the standard pipeline analysis carried out at the {\it Beppo}SAX
Science Data Center.

\subsection{Spectral Analysis}

For the spectral analysis, the LECS data have been considered only in the 
range 0.1--4 keV, due to still some unsolved calibration problems at higher 
energies.
To fit the LECS, MECS, HPGSPC and PDS spectra together, one has to introduce
constant rescaling factors to account for uncertainties in the 
inter--calibration of the instruments. 
The acceptable values for these constants are in the range 0.7--1.0 for the 
LECS and in the range 0.77--0.93 for the PDS, with respect to the MECS. 
The HPGSPC constant is normally in the range 0.9-1 (Fiore et al. 1999). 
The spectral analysis was performed with the XSPEC 10.0 package.

In the fitting procedure we first considered a single power law model
plus absorption with the column density fixed at the Galactic value 
$N_{\rm H}=4.2\times 10^{20}$ cm$^{-2}$.
This model does not give a good fit to the data yielding a 
reduced $\chi^2_r=2.0$ (214 degree of freedom).
The fit improved by allowing $N_{\rm H}$ to vary (see Tab.\,1).
The power law fit gives a good representation
of the 0.1--200 keV X--ray spectrum, with a relative steep photon
spectral index $\alpha = 2.18$. This index is slightly flatter
than the one measured by {\it Rossi}XTE ($\alpha = 2.37 \pm 0.01$
in the 2-10 keV energy band) 
in the same days (1-2 November, Perlman et al 1999). 
Actually, the spectral index determined with {\it Beppo}SAX
is flatter than all spectral indices obtained with
{\it Rossi}XTE in the period October December, 1998.
The flattest {\it Rossi}XTE spectral index is $\alpha = 2.32 \pm 0.015$)
measured on November 6, 1998, a few days before the top
of the X--ray flare (Perlman et al 1999).
Moreover, the very high statistics of the LECS and 
MECS spectra show that from a statistical point of view the power law fit 
is not acceptable and that a more complex model is required. 
This is also reflected by the fact that the interstellar absorption, 
$N_{\rm H}$, determined in the best fit procedure is higher, at 90\% 
confidence, than the Galactic value. This implies either  the
existence of an intrinsic absorption in
PKS 2005--489 which seems unlikely, or that we are using a wrong model.

We therefore considered a broken power law model which gave a much better
fit, even if we keep the $N_{\rm H}$ fixed at the Galactic value.
The difference between the two spectral indices, although significant,
is small, with the break at $\sim 2$ keV. The spectral index above the
break is the same as the single power law, while before the break it
is somewhat flatter (see Tab.\,1 and Fig.\,2). Of course, this break
at 2 keV could not have been detected by {\it Rossi}XTE.
In this fit the LECS, HPGSPC and PDS intercalibration constant
factors have also better values (see notes in Tab.\,1).
Clearly we are detecting a downward curved spectrum,
probably the declining part of the synchrotron emission.
Thus, we considered a more physical intrinsically curved
spectral model developed by Fossati et al (2000b). This model
is an analytic expression that assumes a continuously changing
slope between two asymptotic limits.
It is expressed in a form that provides the following parameters
$(E_1, \alpha_1, E_2, \alpha_2, E_B, f)$, where $E_1$ and $E_2$
are the two ``pivoting energies'', while $E_B$ and $f$ determine
the scale length of the curvature. In order to fit this model to the data
we have to fix one parameter for each of the two pairs
($E_1$, $\alpha_1$) and ($E_2$, $\alpha_2$). If we fix the energy we will
find the spectral index at that energy, while if we fix the spectral
index we will find the energies where it will occur. The latter possibility
is very important, because it allows us to find the peak of the synchrotron
component, if this is inside our energy band, and to estimate its
uncertainty. This can be obtained by setting one spectral index,
for instance $\alpha_1=1$, so that
the best-fit value of $E_1$ will give $E_{peak}$. 
For more details we refer to the Fossati et al. (2000b) paper.
In order to determine the values of the really interesting parameters
and their confidence intervals, in the best-fit procedure we kept
the absorption column density fixed to the Galactic value and $f$
fixed to the best-fit value $f=4.5$.
If we fix the two energies at $E_1=0.1$ and $E_2=100$ keV, we find that
the respective photon spectral indices are $\alpha_1=1.98^{-0.03}_{+0.02}$
and $\alpha_2=2.23 \pm 0.02$. 
We then fixed the photon energy index $\alpha_1=2$ in order to find the
synchrotron
peak and indeed we find that $E_{peak}=1.15^{- n.c.}_{+0.15}$ keV.
This confirms the presence of a curved spectrum with a synchrotron 
peak below $\sim 1.5$ keV. It is in any case remarkable that this curvature
from 0.1 up to 100 keV is rather small with respect to other sources.
All quoted errors, for the curved model, are calculated for
$\Delta \chi^{2}=2.7$ that corresponds to a 90\% 
confidence range for one parameter of interest.

\begin{table*}
\caption{Fit results for a power law and broken power law model}
\begin{center}
\begin{tabular}{|l|cclcccc|}
\hline

model$^a$ & $\Gamma_1^{\rm \ b}$ or  $N_{\rm H}$ & $\Gamma_2$ & break & 
$\chi ^2_r$ (d.o.f.) &$F_{[0.1-2~{\rm keV}]}^{\rm d}$ &$F_{[2-10~{\rm keV}]}$ 
&$F_{[10-100~{\rm keV}]}$ \\
 &  &  &  energy$^{\rm \ c}$ & & erg cm$^{-2}$ s$^{-1}$ & 
erg cm$^{-2}$ s$^{-1}$ & erg cm$^{-2}$ s$^{-1}$ \\
\hline
 & & & & & & & \\
power law & $(4.9^{-0.15}_{+0.15} ) \times 10^{20}$ & $ 2.18^{-0.02}_{+0.02}$ 
                          & & 1.4 (213) & & & \\
broken p. l. &$2.02^{-0.04}_{+0.03}$ &$2.21^{-0.02}_{+0.02}$ 
                          &$1.85^{-0.35}_{+0.45}$ &0.90 (212)       
                          & $4.0\times 10^{-10}$  &  $1.8\times 10^{-10}$
                          & $1.7\times 10^{-10}$  \\

& &  & & & & & \\
\hline
\end{tabular}
\end{center}
$^a$ we had also a constant to allow for uncertainties in the
intercalibration of the instruments (see sec. 2.2); the best fit
values are 1) PL: LECS constant$ = 0.75$, MECS constant$ = 1.$ frozen,
HPGSPC constant$ = 0.94$, PDS constant$ = 0.72$; 2) BPL:  LECS constant$ = 0.78$, 
MECS constant$ = 1.$ frozen, HPGSPC constant$ = 0.97$, PDS constant$ = 0.77$;
$^{\rm b}$ photon spectral index; $^{\rm \ c}$ energy values in keV;
$^{\rm d}$ corrected for the absorption.
\end{table*}

\begin{figure}
\begin{center}
{\resizebox{\hsize}{!}{\includegraphics{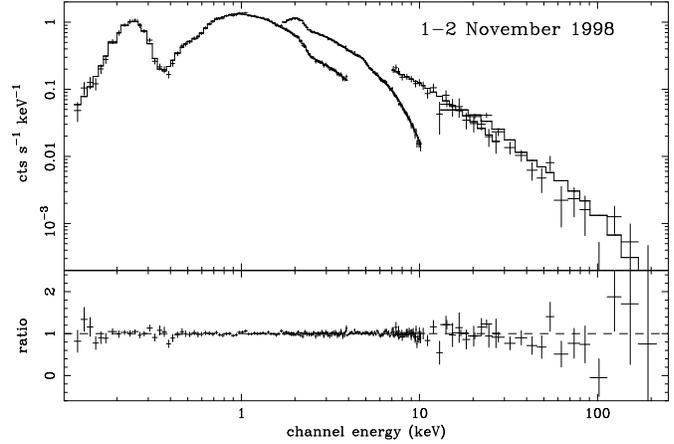}}} \\
\end{center}
\caption{
LECS+MECS+HPGSPC+PDS spectrum observed
on 1--2 November 1998, the best fit model is a broken power law model.
The $N_{\rm H}$ was fixed to the Galactic value.
}
\end{figure}

\section{Infrared Observations}

The $JHKL$ observations were made using the MkIII infrared
photometer attached to the SAAO 1.9m telescope at Sutherland,
in July 1997 and in October 1998, the latter was
one day before the start of the {\it Beppo}SAX observation.
The chopper throw was 30 arcsec in a N-S direction and the 
aperture was 12 arcsec diameter.
This aperture contain a large fraction of the galaxy
whose H-flux is comparable ($\sim 20\% $ lower) to the H-flux
of the nucleus (see Kotilainen et al. 1998). The values reported
in Tab. 2 and Fig.\,4 are not corrected for the galaxy contribution.
However, on the scale of Fig.\,4 this correction is negligible and
smaller than the size of the infrared data symbols.
The calibration in terms of log flux in W m$^{-2}$ Hz$^{-1}$ 
for a zeroth mag star on this system at $JHKL$ are -22.82,
-23.01, -23.21 and -23.55 respectively.

As it is usually found for HBL, there is very little change in the
slope of the infrared spectrum as the luminosity changes (see Tab.\,2),
i.e. the infrared colour of the blazar changes little or not at all.
This is due to the fact that the synchrotron peak is at much higher 
energies. Changes in $K$ of up to four mags have been seen in some
blazars with little change in their infrared colours
(e.g. Fan 1999; Hagen-Thorn et al. 1994).
In the particular case of PKS\,2005--489, the $K$ mag changed only by
0.32 mag during the observations, which is relatively little for
this class of objects.

\begin{table}
\caption{Infrared Observations}
\begin{tabular}{llcccc}
\hline
 & & & & &\\
date     &Jul. Date   &J$^{\rm a}$      &H       &K    &L   \\
27-07-97 &2450657     &10.97  &10.23   &9.55 &8.52$\pm$0.03 \\
         &2450721     &11.17  &10.41   &9.75 &8.80$\pm$0.08 \\
         &2450738     &11.23  &10.51   &9.87 &8.91$\pm$0.05 \\
         &2450742     &11.26  &10.49   &9.84 &8.70$\pm$0.07 \\
30-10-98 &2451117.3   &11.16  &10.46   &9.80 &8.69$\pm$0.04 \\
\hline
\end{tabular}

$^{\rm a}$ the errors for the J, H, K magnitudes are $\le 0.03$.

\end{table}

\section{Discussion}

\subsection{SED}

We observed PKS 2005--489 while it was in a very high state, aiming 
at detecting a change in its spectral energy distribution (SED) 
with respect to previous observations.
Very significant changes of the overall SED were in fact detected in 
other HBL objects, such as MKN 501 (Pian et al. 1998) and 1ES 2344+514 
(Giommi et al 2000): in these objects, during flares,
the synchrotron peak frequency increased by 2--3 orders of magnitudes 
with respect to quiescent states.
These observations provide new conditions
that the various models have to reproduce, exploring new range of values
for the physical parameters and allowing us to investigate
the accelerating mechanisms inside the jets.

In the case of PKS 2005--489, the X--ray data can be fitted equally
well by a broken power law and by a continuously steepening model. 
Although the spectral parameters are not very different, the
two models predict a quite different frequency for the
peak of the synchrotron emission, as illustrated in the two panels
of Fig. 4. In the case of a broken power law in fact, it can be possible
to interpret the entire far--IR to hard X--ray SED as due
to a single synchrotron component, peaking at $\sim 10^{15}$Hz.
If, instead, the X--ray data are fitted by the continuously steepening
model, then the synchrotron peak is at $\sim 0.1$ keV, and the IR flux
must be produced by another jet component.
This uncertainty in locating the synchrotron peak frequency 
originated the two possible sets of parameters 
of the homogeneous synchrotron self Compton model
by which we can interpret the overall SED, discussed below.

In any case, our data show that the source was in a very high state 
and that the 0.1--200 keV spectrum is well fitted by a spectrum
due to synchrotron emission only. 
In Fig. 3 we plot an enlargement of the SED in the X--ray spectral region. 
It can be seen that PKS 2005--489 underwent dramatic flux changes,
but, at least above $\sim$1 keV, the spectral shape was always
steeper than unity, constraining the peak of the synchrotron emission
to be at lower energies.
Our X--ray spectrum is the flattest ever observed,
even considering the many observations by {\it Rossi}XTE during
the same period of our 1998 {\it Beppo}SAX observation
(Perlman et al. 1999).

\begin{figure}
{\resizebox{\hsize}{!}{\includegraphics{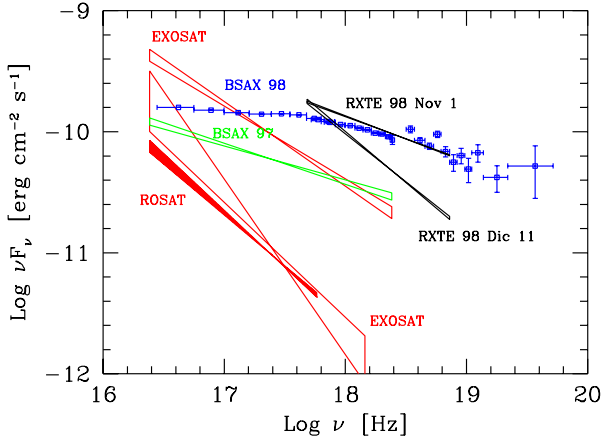}}}\\
\vskip -2.5 true cm
\caption{
The X--ray energy spectra of PKS 2005--489 as observed at different
epochs by different satellites. 
Source of data: 
Sambruna et al 1994b (EXOSAT),
Comastri et al. 1997 (ROSAT),
Perlman et al. 1999 ({\it Rossi}XTE).
}
\end{figure}

\begin{figure*}
{\resizebox{\hsize}{!}{\includegraphics{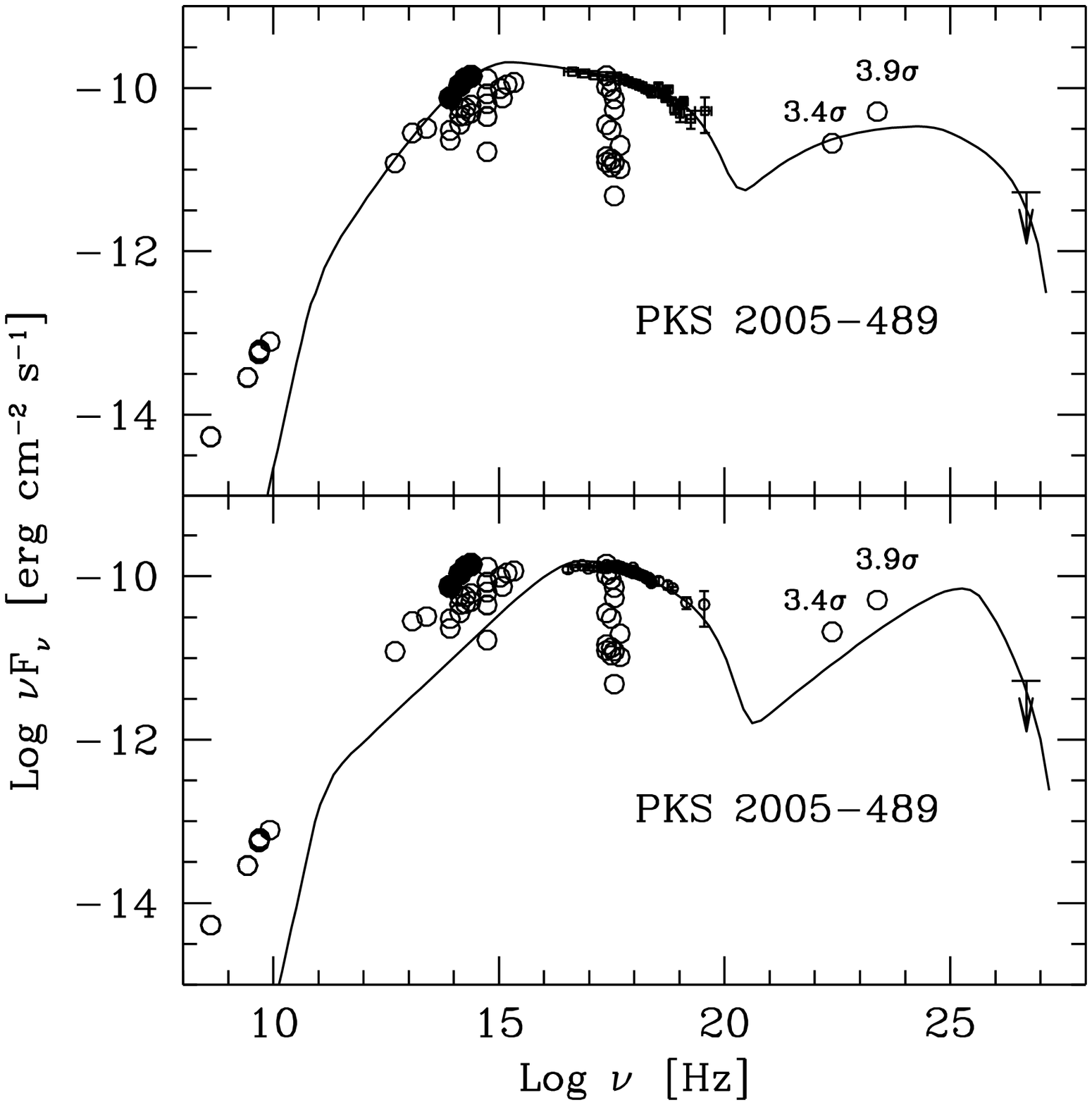}}}\\
\caption{The SED of PKS 2005--489.
Note that only the {\it Beppo}SAX and the IR data (filled symbols)
are simultaneous.
The other data have been taken from the Nasa Extragalactic Database (NED).
In the top panel we used for the X--ray spectrum the broken 
power law best--fit model. 
In this case the SSC model fits also the IR data. 
In the bottom panel we used for the X--ray spectrum the curved 
best--fit model (see text).
In this case the IR emission is not due to the same SSC component
responsible for the X--ray emission.
For both SSC models we have assumed no external photons contributing
to the inverse Compton process.
The input parameters for the top (bottom) models are:
$L^\prime_{inj}= 5.5\times 10^{41}$ ($7.8\times 10^{41}$)  erg s$^{-1}$,
$B=1$ (1) Gauss, 
$R=1.5 \times 10^{16}$ $ (10^{16}$) cm, 
$\gamma_{min}= 3\times 10^3$ ($1.5 \times 10^{4}$),
$\gamma_{max}= 10^{6}$ ($1.25\times 10^6$), 
$s=2.2$ (2.3),
$\delta = 18$ (15).
}
\end{figure*}

\subsection{Limits on magnetic field and particle energies}

The spectral properties of our X--ray data indicate that
the dominating radiation process is synchrotron.
Therefore the emission is due to the most energetic electrons,
suffering rapid radiation energy losses.
During our observation the flux variability was modest (10\%
on a timescale of about 10 hours), 
but EXOSAT, in two occasions, observed flux changes of a factor 
of about 4 in 4--5 hours (Giommi et al. 1990).
We can then require that the emitting electrons cool radiatively 
(by synchrotron and inverse Compton) on these timescales,
implying a lower limit on the magnetic field. 
Assuming equal radiation and magnetic energy densities,
using $t_{\rm var}=4$ hours and $\nu_x=1\times 10^{17}$ Hz, we derive 
$B>0.3\delta^{-1/3}$ Gauss and $\gamma_x< 3\times 10^5\delta^{-1/3}$.
Here $\delta=[\Gamma-\sqrt{\Gamma^2-1}\cos\theta]^{-1}$ 
is the Doppler beaming factor, where $\theta$ is the viewing angle.

\subsection{Homogeneous SSC model}

We have fitted the SED of PKS 2005--489 with a pure, homogeneous and
one zone SSC model, assuming no contribution from the photons produced 
externally to the jet as seeds for the Compton scattering process.
This class of SSC models is completely constrained once the peak
energies and their fluxes are known.
Unfortunately, in our case only the X--ray and IR data are simultaneous,
and the location of both peaks (particularly the Compton one),
is somewhat uncertain.
However, the X--ray data alone constrain the synchrotron
peak to be at or below the low energy end of the BeppoSAX range
(i.e. $\le 0.1$ keV).
For the high energy part of the SED we have considered the previous
observation by EGRET, resulting in a marginal detection, see Fig. 4,
(Lin et al. 1997; von Montigny et al. 1995) and the TeV upper limit
given by CANGAROO (Roberts et al. 1999). These data constrain the
Compton peak to be close to $10^{25}$ Hz.

Given the uncertainty in the location of the synchrotron
peak, we have considered two models, shown in the two panels
of Fig. 4.
The top panel shows the X--ray data fitted by a broken power law
model, and in this case it is possible that the simultaneous IR flux is 
produced in the same region producing the high energy emission.
We have then considered a source with size $R=1.5\times 10^{16}$ cm
and a beaming factor $\delta=18$, resulting in a minimum observed 
variability timescale of $\sim$6 hours, which is typical for this 
class of sources, and close to the minimum variability timescale 
observed in this source by EXOSAT (Giommi et al. 1990).
With these values of the size and beaming factor, the magnetic
field is constrained to be of the order of $B=1$ Gauss, in order
to have nearly equal synchrotron and self--Compton luminosities,
as observed.
The intrinsic power (assumed to be continuously injected throughout 
the source in the form of relativistic electrons) is required to be 
$L_{inj}=5.5\times 10^{41}$ erg s$^{-1}$.
The relativistic electrons are assumed to be injected with a
power law energy distribution of slope $s=2.2$ between
$\gamma_{min}=3\times 10^3$ and $\gamma_{max}=10^6$.
The value $s=2.2$ is required to match the slope of the X--ray spectrum.
The resulting SSC spectrum is shown as a solid line in the top panel
of Fig. 4.
It can be seen that this model also agrees with the 
(admittedly uncertain) flux in the EGRET band,
and well describes the entire SED
(except for the mm--radio part, which should come from larger
jet regions, less affected by synchrotron self--absorption
at these frequencies).
Being produced by the same region, we expect some variability
correlations between the IR and X--ray flux.
In particular, if the radiative cooling time {\it at both frequencies}
is shorter than the light crossing time, then it is the latter
timescale that is observed in both energy bands.
With the adopted parameters, this is indeed the case:
at the observed frequency $\nu=3\times 10^{14}$ Hz, 
the intrinsic cooling time is half the crossing time.

The bottom panel of Fig. 4 shows the X--ray data fitted by the
continuously steepening model.
In this case the IR flux cannot be produced by the same 
region emitting the X--rays, resulting in a significantly
larger frequency of the synchrotron peak.
The adopted parameters of the SSC model are similar to
the previous case (see the caption of Fig. 4).
The only significant difference is the value of 
$\gamma_{min}=1.5\times 10^4$, now much larger.

In this case we predict/expect that the IR flux has a different 
variability behavior than the X--ray flux, since it would probably
be produced in a larger region of the jet, with a smaller 
value of the magnetic field, making the cooling time
of the IR emitting electrons longer.
Furthermore, if the variability is caused by a perturbation
of a steady jet propagating from the inner to the outer regions,
we expect some delay between the X--ray and the IR flux
variations, giving information about the relative distance
of the two regions.

\acknowledgements{
This research was financially supported by the Italian
Space Agency. We thank the {\it Beppo}SAX Science Data
Center (SDC) for their support in the data analysis.
This research made use of the NASA/IPAC Etragalactic Database (NED)
which is operated by the Jet Propulsion Laboratory, Caltech, under
contract with the National Aeronautics and Space Administration.
}

\end{document}